\documentstyle[12pt]{article}
\newlength{\aivwidth}   
\newlength{\tmpwidth}   

\title{ON GRAVITATION AND QUANTA\thanks{This is a slightly extended
version (with a footnote added on December 19, 1997) of an
Abstract submitted in March 1997 to the Eight Marcel Grossmann Meeting,
Jerusalem, June 22-27, 1997.}}

\author{Pawel O. Mazur  \\
        Department of Physics and Astronomy, \\
        University of South Carolina, Columbia, SC 29208, USA
        \thanks{E-mail address: mazur@psc.psc.sc.edu}}
\date{January 5, 1997}
\begin{document}
\maketitle

\newcommand{\be}{\begin{equation}\label}
\newcommand{\ee}{\end{equation}}
\newcommand{\ba}{\begin{eqnarray}\label}
\newcommand{\ea}{\end{eqnarray}}
\newcommand{\pl}{\partial}
\newcommand{\lb}{\lambda}
\newcommand{\kp}{\kappa}
\newcommand{\tr}{\triangle}
\newcommand{\dt}{\delta}
\newcommand{\bt}{\beta}
\newcommand{\al}{\alpha}
\newcommand{\gm}{\gamma}
\begin{abstract}

 {We show that a gravitating mass $M$ in thermal equilibrium behaves
statistically like a system of some number $N$ of harmonic oscillators
whose zero-point energy ${\epsilon\over 2}$ depends on $N$ universally
in such a way that $N{\epsilon}^2\sim {{hc^5}\over G}$,
where $G$ is the Newton constant, $c$ is the velocity of light
in vacuum, and $h$ is the Planck constant.
The large number $N$ is also the number of weakly interacting
gravitational atoms [2] which are the constituents of a black hole.
The sum over all oscillators of the squares of zero-point energies
is fixed and independent of the number of those oscillators.
It is well known that the classical gravitating systems behave
in the way foreign to statistical quantum mechanics.
The negative specific heat of those systems and the phenomenon
of a gravitational collapse are different facets of the same reality.
The enigmatic Bekenstein entropy of black holes was not yet derived
on the basis of microscopic theory. Our work may be considered a
first step in direction of presenting such a basis [1-4].
The problem with all present approaches to this problem has been the
silent assumption that the total entropy of gravitating systems
(black holes) {\it must} be given by the Bekenstein formula
$S_{bh} = 4k{\pi}M^2$, where the mass $M$ is in Planck units.
The postulate of gravitational constituents (gravitational
atoms) and gravitational oscillators (quanta)
leads to Bekenstein formula only after a part of mass-energy
fluctuations is neglected [2].
The most unusual character of the gravitational mass-energy
oscillators (quanta) is that they somehow manage, via the quadratic
sum rule defining the Newton constant $G$,
$\sum_i {\epsilon_i}^2 = b{hc^5 \over G}$,
to reduce their zero-point energy when
the number $N$ of gravitational atoms grows [2].
This also means that a cold
large gravitational mass $M\sim \mu{\sqrt N}$ consists of $N$
{\it constituents}. The formula $M^2 ={{\mu^2}\over 2\pi}N$,
${\mu}^2 = {hc\over G}$, was derived long time ago by
the present author. The physical meaning of the
`phenomenological' entropy of Bekenstein is that
it is the measure of the number $N$ of {\it constituents}
making up a very `cold' large body. The zero-point energy $\epsilon_i$
of {\it gravitational quanta} for a very large `cold' mass is of an
order of the Hawking thermal energy of quanta,
${\epsilon_i}\sim {{\mu^2}\over (4{\pi})^2 M}$. The more massive
is a gravitating mass the softer are the {\it gravitational quanta}.
The number $N$ of {\it constituent gravitational atoms} of a
given spin-zero quantum Schwarzschild black hole determine
the energy of quasi-thermal quanta and the Bekenstein-Hawking
entropy, $kT_{bh}\sim {\mu\over \sqrt{N}}$, $S_{bh}\sim kN$, but it
is valid only in the particular limit when the interference terms are
neglected [2]. Otherwise, as usual
with oscillators, there are two sources of statistical fluctuations
of mass-energy corresponding to the corpuscular and wave
aspect of quanta [2]. We calculate the energy fluctuations of
the system considered in [2]\footnote{Two examples of gravitating
systems are discussed in
[2] simultaneously: a spin-zero quantum black hole and the whole
non-rotating Universe. The present upper bound on cosmological constant
$\lambda$ in Planck units,
regarded as the vacuum zero-point energy density,
was used to estimate the lower bound on
the total number $N_U\sim 10^{123}$ of gravitational atoms
(and gravitational oscillators) in the Universe.}
in terms of the mean energy
$\overline{E}$, where
$\overline{E} = {E_0}cotanh({\epsilon\over {2kT}})$, $\epsilon =
{E_0\over N}$, ${E_0}^2 = {b{\mu}^2{c^4}N\over 4}$, and
$\overline{(\Delta E)^2} = N^{-1}{\overline{E}}^2 - {1 \over
4}N{\epsilon}^2 = N^{-1}{\overline{E}}^2 - {1 \over 4}b{\mu}^2c^4 .$
Neglecting the ${1 \over N}$ term in this formula,
when $\overline{E}$ is fixed,
we obtain the expression for statistical fluctuations typical
of gravitating systems:
$\overline{{(\Delta E)^2}_{bh}} = -{1\over 4}b{\mu}^2c^4 .$
The well known relation between the mass-energy fluctuations
and the behaviour of entropy near the state of thermal equilibrium,
$\overline{(\Delta E)^2} = - k\biggl({\partial^2 S \over \partial E^2}
\biggr)^{-1} ,$ leads to the entropy of such a {\it truncated system}:
${\partial^2 S_{bh} \over \partial E^2} = 4kb^{-1}{\mu}^{-2}c^{-4}$.
Integrating this last equation gives the inverse temperature
$\beta_{bh} = 4b^{-1}\mu^{-2}c^{-4}{\overline{E}}$, where an arbitrary
integration
constant is fixed to be zero by demanding that a very massive body
is also very cold [2]. The entropy is given by the
`phenomenological' entropy formula of Bekenstein:
$S_{bh} = 2kb^{-1}{\mu}^{-2}{\overline{E}}^2 + const$.
The model calculation of Hawking leads to a numerical
value of the constant $b$, $b = {1 \over 4{\pi}^2}$.
Quite independently of the actual value of the numerical constant $b$
the entropy $S_{bh}$ has a lower bound
$S_0 = 2kb^{-1}{\mu}^{-2}{E_0}^2$ which depends only on $N$.
This follows from the fact that the total mass-energy $\overline{E}$ is
bounded from below by the zero temperature value $E_0$,
${\overline{E}}\geq E_0$. Now, ${E_0}^2 = {1\over 4}b\mu^2c^4N$, and,
therefore, the lower bound on the entropy does not depend on $b$ at all,
$S_0 = k{N \over 2}$. It is quite natural for the entropy to be
bounded from below by the number ${N \over 2}$ of {\it
constituents}\footnote{It is clear to this author that only the
large $N$ limit (${1\over N} = \nu\rightarrow 0$, where $\nu$ plays the
role analogous to the Planck constant in the new gravitational
noncommutative mechanics [1-4,8,9]), with fixed observable
quantities like $\overline{E}$, produces such an expression for
entropy. If we were to impose the Boltzmann statistics on the very
`cold' collection of gravitational atoms this would have been so much
against the spirit and the letter of quantum theory
that we would have been led to the conclusion that
`quantum gravity' indeed requires really revolutionary ideas
like the abandonement of {\it quantum statistics}.
Such revolutionary step forward would make it possible
to reproduce the Bekenstein-Hawking entropy
by an order of magnitude estimate. On the other hand we are not
prepared for such a revolutionary step forward yet.

The analogy which invites itself quite naturally is this:
in the limit $\hbar\rightarrow 0$ in quantum statistical mechanics of
ideal gases when we neglect the fact of identical nature of quantum
particles we obtain the classical Boltzmann gas plagued with its
Gibbs paradox [6] and etc..
We should perhaps recall here the story of the now well understood
phenomenon of superfluidity in the liquid Helium II. The moral of this
very well known
story, as told by R. P. Feynman [7], is this: {\it DO NOT QUANTIZE WHAT
IS ALREADY QUANTIZED}. This basic observation [7] which was found valid
for superfluidity in the past is equally valid today for quantum black
holes [2,4,8,9].
Landau's {\it quantum hydrodynamics} was only formally `quantum' in
view of quantum commutators appearing
in the nonrelativistic current algebra.
Feynman has discovered [7] that the real problem with {\it quantum
hydrodynamics} was that hydrodynamical description of the superfluid
Helium II failed to take into account
the Bose statistics of Helium 4 atoms.
Now, it is our opinion that we seem to be facing the
same dilemma posed by the currently fashionable
description of black holes.
Incidentally, the cases for quantization of general relativity (GRT)
and its later fermionic deformation known as `supergravity'
seem to follow the same pattern which was so well understood by Feynman
[7] in the context of experimentally rich phenomenon
of superfluidity. It seems that general relativity, `supergravity',
`superstrings' and recently `supermembranes' were also quantized
in the same way the superfluid hydrodynamics was quantized
with the well known results [7].
Again, the moral of the story as told
by R. P. Feynman [7] when adopted to the present situation of quantum
black holes seems to be that ``we should give to the emperor what is
emperor's''. The Atomic Hypothesis and Quantum Statistics rule.
Gravitation is the many-body phenomenon [9]. It is our
responsibility now
to find the physically correct Hamiltonians for systems of gravitational
atoms in the framework
of the new gravitational noncommutative mechanics [1,3,4,8,9].}.

We have seen the emergence of the Bekenstein formula for the
black hole entropy from the hypothesis about the microscopic nature of
gravitational phenomena [1,2,4,8,9]. We shall propose to apply the same
argument to the whole Universe. It seems natural to consider $N_U\sim
10^{123}$ gravitational atoms [2] and apply to them the same physical
argument. We obtain the simple estimate [2] for cosmological constant
regarded as the zero-point vacuum energy density ${\lambda}\sim
{\mu^4\over N_U}$. It was shown [8] that a system of a large
number of gravitational atoms described in the framework
of the Atomic Theory of Gravitation (the new gravitational mechanics)
by a large $N\sim N_U$ matrix of operators [1,2,8,9] bears remarkable
similarity to the thermal Universe with a nonvanishing positive
cosmological constant.

If the Bekenstein entropy were the whole thing as
far as the thermal properties of gravitating masses are concerned, then
the World would be always in a state of the lowest thermodynamic
probability. This conclusion would lead then to the statement that
the behaviour of a visible Universe is determined by the condition
that it is in a state of the lowest statistical weight.
Considering an ensemble of such Universes, regarded as local
thermal phenomena in a sense suggested in [2,4,8], we would
be persuaded to conclude that our Universe is the least probable one.
The Universe must be regarded as a very typical one in the
{\it statistical ensemble of Universes}, which is also the statement
of the maximal thermodynamic probability $W$ of Boltzmann.
It should be noticed that the notion of a statistical ensemble
for the observable Universe is justified only after we identify
{\it atoms} whose existence is underlying the totality of phenomena.
The Gibbs-Jaynes principle of the maximum of $H$ function [5] is
applicable to a closed system once we postulate the general
{\it Atomic Hypothesis} [1-4].
According to this hypothesis the totality of phenomena
should be derived from the properties of space-time-matter
atoms, which I prefer to call {\it gravitational atoms} [1-4].
The result described in this short note ({\it Abstract}), which
was based
on the purely phenomenological considerations, was first derived in the
Spring of 1995 and published in [2]. I have applied the simple
interpolation argument, originally due to Planck, to the problem of
mass-energy fluctuations of a gravitating mass. I have proposed that
the formalism of {\bf the new gravitational noncommutative mechanics}
[1-4,8,9] be applied to gravitating particles (black holes)
and to the whole Universe.

The point made very clearly in [1] was that Hamilton's optical analogy
should be taken seriously for both GRT and QM.
It is obvious to the present author that this analogy is quite useful
once we realize that the perihelion precession of the Mercury, and
the {\it Complementarity Principle} for a gravitational mass and a
inertial mass aspect of a gravitating particle [1,3,8,9] allow for the
transition to new gravitational mechanics
in the same way the nonrelativistic Kepler
problem and the spectroscopic data has made it possible for Heisenberg
to propose the matrix quantum mechanics [1]. The idea of using the GRT
Kepler problem and the `optical analogy' of Hamilton to make a similar
transition to the new gravitational noncommutative mechanics was first
described by the present author [1,3] as early as in the Spring 1995
(the submittal date of [1] was May 31, 1995),
and later during the USC Summer 1995 Institute [3]. The idea of what
I have called the {\it Second Heisenberg Algebra}
and the {\it Second Period of Nature} was described in numerous lectures
and talks (CALTECH, February 1996; Tel-Aviv University, March 1996;
Ecole Polytechnique, Paris, October 1996).

The picture of a gravitating particle
(a black hole) which has emerged from my work is not unlike that of
a giant nucleus or a baryon\footnote{The large $N$ QCD description of
baryons and the Wigner (or the modern matrix models)
description of energy levels in nuclei share some common traits.}.
The calculation of specific properties of such a complex object
like a quantum black hole must be carried out in a sort of
large $N$ approximation in the new gravitational noncommutative
mechanics proposed in [1,3,8,9], and it is under way [4].

\item {\it References}

\item {1.} P. O. Mazur, Acta Phys. Polon. {\bf B26}, 1685 (1995);
hep-th{/9602044}.
\item {2.} P. O. Mazur, Acta Phys. Polon. {\bf B27}, 1849 (1996);
hep-th{/9603014}.
\item {3.} P. O. Mazur, Lecture at the Summer Institute at USC, 1995.
\item {4.} P. O. Mazur, in preparation.
\item {5.} E. T. Jaynes, Phys. Rev. {\bf 106}, 620 (1957); ibid. {\bf
108}, 171 (1957);
\item {  } S. P. Heims and E. T. Jaynes, Rev. Mod. Phys. {\bf 34}, 143
(1962);
\item {6.} J. W. Gibbs, {\it Elementary Principles in Statistical
Mechanics}, Yale University Press, 1902.
\item {7.} R. P. Feynman, {\it Application of Quantum Mechanics to
Liquid Helium}, in {\it Low Temperature Physics} {\bf 2}, 17-53 (1955).
\item {8.} P. O. Mazur, {\it On the Fundamental Idea of the New
Gravitational Mechanics},  Spring 1996, unpublished.
\item {9.} P. O. Mazur, {\it Gravitation as a Many Body Problem}, in
{\it Beyond the Standard Model V}, eds. G. Eigen, P. Osland, and B.
Stugu, AIP Proceedings 415, pp. 299-305, New York 1997;
hep-th{/9708133}.}
\end{abstract}
\vskip .5cm
\centerline{\bf Acknowledgements}
\vskip .2cm
This research was partially supported by a NSF grant to USC.
I wish to thank organizers of the Eight Marcel Grossmann Meeting in
Jerusalem, June 1997, for inviting me to this Meeting, which, however, I
was unable to attend.
\end{document}